\documentclass[reprint,superscriptaddress,amsmath,amssymb,aps,prl]{revtex4-2}
\usepackage{graphicx}
\usepackage{braket}

\begin{document}

\title{Charge-Transfer Energy Governed by $f$--$d$ Coulomb Interaction in Rare-Earth Systems}

\author{Norimasa Sasabe}
\email{SASABE.Norimasa@nims.go.jp}
\affiliation{Center for Basic Research on Materials, National Institute for Materials Science, Tsukuba, Ibaraki 305-0047, Japan}

\author{Shinji Watanabe}
\affiliation{Department of Basic Sciences, Kyushu Institute of Technology, Kitakyushu, Fukuoka 804-8550, Japan}

\author{Akira Yasui}
\affiliation{Japan Synchrotron Radiation Research Institute, Sayo, Hyogo 679-5198, Japan}

\author{Naomi Kawamura}
\affiliation{Japan Synchrotron Radiation Research Institute, Sayo, Hyogo 679-5198, Japan}

\author{Kojiro Mimura}
\affiliation{Department of Physics and Electronics, Osaka Metropolitan University, Sakai, Osaka 599-8531, Japan}

\author{Takayuki Uozumi}
\affiliation{Department of Physics and Electronics, Osaka Metropolitan University, Sakai, Osaka 599-8531, Japan}

\author{Masaichiro Mizumaki}
\affiliation{Faculty of Science, Kumamoto University, Chuou-ku Kurokami, Kumamoto 860-8555, Japan}

\author{Hitoshi Sato}
\email{jinjin@hiroshima-u.ac.jp}
\affiliation{Research Institute for Synchrotron Radiation Science, Hiroshima University, Higashi-Hiroshima, Hiroshima 739-0046, Japan}

\date{\today}

\begin{abstract}
The $f$--$d$ Coulomb interaction ($U_{\rm fd}$) plays a central role in low-energy theories describing valence fluctuations and quantum valence criticality in rare-earth systems. In contrast, core-level spectroscopies have conventionally been analyzed using the phenomenological charge-transfer energy ($\Delta$), and the relationship between these two descriptions has remained unclear. Here, we develop a spectroscopic framework demonstrating that the charge-transfer energy contains an $f$--$d$ Coulomb contribution $U_{\rm fd}n_d$ arising from the $d$-electron occupation ($n_d$), leading to the charge-transfer energy $\Delta'$. We apply this framework to the prototypical valence-transition compound YbInCu$_4$ and demonstrate, from combined analyses of core-level photoemission and $L$-edge x-ray absorption spectroscopy, that the observed spectroscopic evolution across the first-order valence transition is consistently interpreted in terms of the variation of $\Delta'$.
\end{abstract}

\maketitle

Rare-earth compounds exhibit a wide variety of emergent phenomena driven by strong electronic correlations between localized $4f$ electrons and itinerant conduction states \cite{Hewson1993book, QCP2001RMP, QCP2006RMP, QCP2007RMP}. Among the local interactions inherent to these systems, the Coulomb repulsion between the rare-earth 4$f$ and 5$d$ electrons, $U_{\rm fd}$, has been identified as a key ingredient governing valence transitions \cite{Falicov1969PRL, Watanabe2009JPSJ, Watanabe2018IOP} and quantum valence criticality \cite{Watanabe2010PRL} in lattice models such as the Falicov--Kimball model \cite{Falicov1969PRL} and the extended periodic Anderson model \cite{Watanabe2009JPSJ, Watanabe2010PRL, Watanabe2018IOP}. Recent advances in core-level x-ray spectroscopies have provided clear experimental signatures of $U_{\rm fd}$ in rare-earth compounds, demonstrating that its magnitude reaches several electronvolts \cite{Kotani1987JPSJ, Ogasawara2000PRB, Tosti2004PRB, Tonai2017JPSJ}. This energy scale exceeds typical crystal-field splittings and spin--orbit interaction energies, indicating that $U_{\rm fd}$ can play a dominant role in shaping the electronic structure.

In theoretical descriptions of core-level spectroscopies, the effect of $U_{\rm fd}$ has been extensively studied primarily in the final state \cite{Kotani1987JPSJ, Tonai2017JPSJ, Kotani2015JPSJ}, where the creation of a core hole strongly perturbs the local electronic environment. These approaches successfully explain characteristic spectral line shapes and satellite structures observed experimentally. By contrast, the role of $U_{\rm fd}$ in the initial state---prior to core-hole creation---remains largely unexplored within microscopic impurity-based frameworks, leaving the connection between spectroscopic observables and low-energy electronic properties incomplete. Although the initial-state electronic configuration contains essential information on valence, charge fluctuations, and their evolution across phase transitions, parameters extracted from spectroscopy, such as the charge-transfer energy $\Delta$ and the $c$--$f$ hybridization strength $V_{\rm cf}$ \cite{Kotani1973JPSJ, GS1983PRB, ZaanenPRL1985}, have conventionally been treated phenomenologically without explicitly incorporating the contribution of $U_{\rm fd}$. As a consequence, the microscopic origin of spectroscopically determined parameters has remained unclear, obscuring the direct correspondence between core-level spectroscopy and valence physics.

To resolve this issue, we propose a theoretical framework for core-level spectroscopy that explicitly incorporates the $f$--$d$ Coulomb interaction $U_{\rm fd}$ into the charge-transfer energy already in the initial state.
This formulation reveals the microscopic contribution of $U_{\rm fd}$ to the charge-transfer energy, providing a direct connection between x-ray spectroscopic observables and low-energy electronic properties.
Within this framework, initial- and final-state effects are treated on equal footing, enabling direct access to microscopic information on valence and charge fluctuations. As a concrete application, we investigate the mixed-valence compound
YbInCu$_4$ \cite{YbInCu4PRB1986, YbInCu4PRB1994, YbInCu4PRB1996}, which exhibits a first-order valence transition, and show
that incorporating the initial-state contribution of $U_{\rm fd}$ enables
a unified interpretation of the spectroscopic data. Our results
indicate that the spectroscopically determined charge-transfer energy
can be microscopically related to the $f$--$d$ Coulomb interaction
through the $d$-electron occupation, providing a microscopic framework
that connects core-level spectroscopy with low-energy valence physics
in rare-earth compounds.

\begin{figure}[t]
\includegraphics[width=8.2cm]{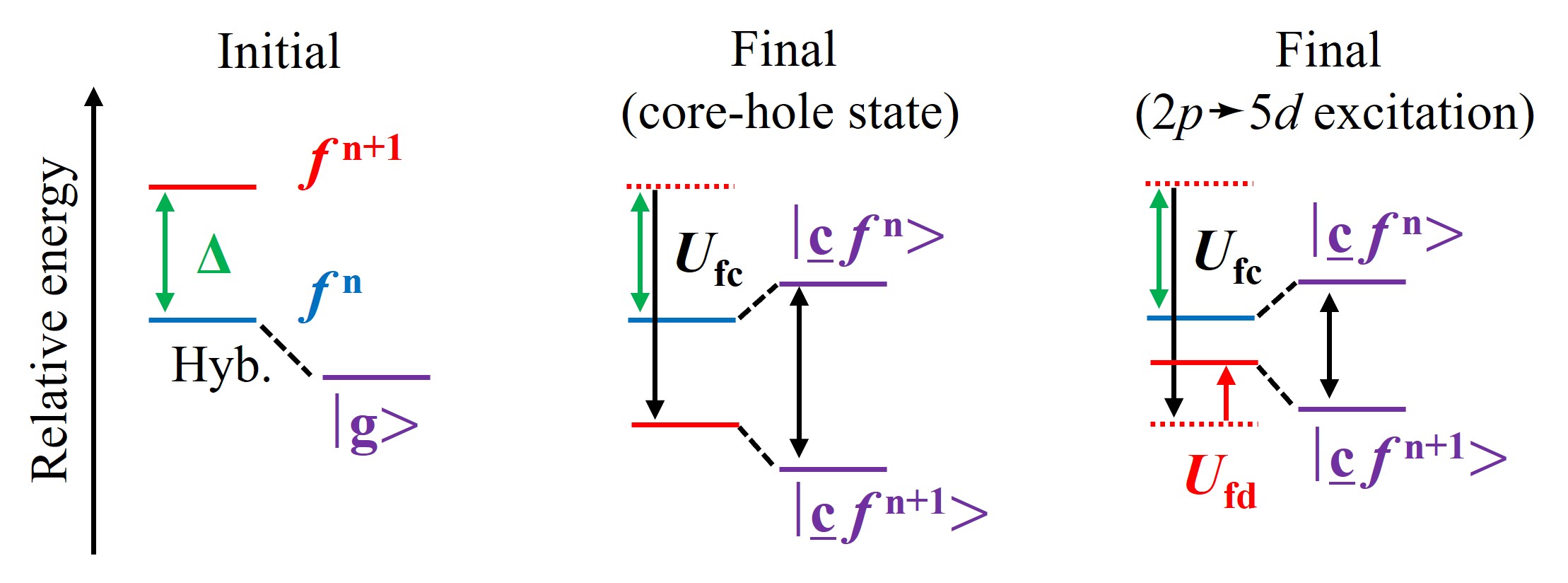}
\caption{
Schematic energy-level diagrams of the initial and final states in core-level spectroscopies. The central panel represents the core-hole final states reached by photoemission into the vacuum level. The right panel shows the final state following the $2p \rightarrow 5d$ excitation. Here, $\underline{\rm c}$ denotes a core hole, while $\Delta$, $U_{\rm fc}$, and $U_{\rm fd}$ denote the charge-transfer energy, the Coulomb interaction between the core hole and the $4f$ electrons, and the Coulomb interaction between the $4f$ and $5d$ electrons, respectively.
}
\end{figure}

Figure~1 schematically illustrates the theoretical framework employed in this study \cite{GS1983PRB, GS1983PRL, Kotani2008book}. The left panel represents the initial state of a rare-earth ion, where the ground state $|g\rangle$ is formed by hybridization between the $f^{n}$ and $f^{n+1}$ configurations.
The charge-transfer energy is defined as the energy difference between these configurations, while their mixing is governed by the $c$--$f$ hybridization strength.
The central and right panels show representative final states relevant to core-level spectroscopies. The central panel corresponds to a core-hole final state, where the core-hole potential $U_{\rm fc}$ strongly perturbs the local electronic structure. Such final-state effects constitute the conventional framework of core-level spectroscopy. By contrast, the right panel describes the $2p \rightarrow 5d$ excitation process, where the $5d$ states explicitly participate in the final state \cite{Kotani2011JPSJ, Kotani2015JPSJ}. In this case, the spectral structure is affected not only by $U_{\rm fc}$ but also by the $f$--$d$ Coulomb interaction $U_{\rm fd}$.

The charge-transfer energy $\Delta$ governs both the initial and final states and can be experimentally determined from spectroscopy \cite{Kotani2008book}. In the conventional formulation, the configuration-averaged energies of the $f^{n}$ and $f^{n+1}\underline{L}$ configurations are written as
\[
\left\{
\begin{aligned}
E_{\mathrm{av}}^{(0)}(f^{n})
&=
n\varepsilon_f
+\frac{n(n-1)}{2}U_{\rm ff}
+E_a, \\
E_{\mathrm{av}}^{(0)}(f^{n+1}\underline{L})
&=
(n+1)\varepsilon_f
+\frac{n(n+1)}{2}U_{\rm ff}
-\varepsilon_L
+E_a,
\end{aligned}
\right.
\]
where $\underline{L}$ denotes a ligand hole, $\varepsilon_f$ is the $f$-level energy, $U_{\rm ff}$ the intra-atomic Coulomb interaction, $\varepsilon_L$ the ligand level, and $E_a$ a reference energy. These parameters correspond to microscopic quantities appearing in Anderson-type lattice models for strongly correlated systems \cite{Anderson1961PR}, whereas the charge-transfer energy is introduced as a phenomenological parameter incorporating their combined effects \cite{ZaanenPRL1985}. The conventional charge-transfer energy is therefore defined as
\begin{equation}
\Delta^{(0)}
\equiv
E_{\mathrm{av}}^{(0)}(f^{n+1}\underline{L})
-
E_{\mathrm{av}}^{(0)}(f^{n})
=
\varepsilon_f+nU_{\rm ff}-\varepsilon_L.
\label{eq:Delta}
\end{equation}

Although Eq.~(\ref{eq:Delta}) has been widely used in the analysis of core-level spectroscopies \cite{Ufc1992JPSJ, Kotani2001RMP, Sasabe2023JPSJ}, it does not explicitly include the contribution of the $f$--$d$ Coulomb interaction in the initial state. Consequently, the charge-transfer energy remains a phenomenological quantity, and its microscopic connection to the key interaction driving valence instability in the Falicov--Kimball and extended periodic Anderson models remains obscure.

To overcome this limitation, we explicitly incorporate the initial-state $f$--$d$ Coulomb interaction through the rare-earth $4f$ and $5d$ occupations, $n_f$ and $n_d$, respectively. Here, $n_d$ denotes the $5d$ occupation prior to core-hole creation \cite{YasuiPRB2011, YasuiPRB2013}. The average energies are then rewritten as
\begin{equation}
E_{\mathrm{av}}(f^{n})
=
E_{\mathrm{av}}^{(0)}(f^{n})
+
U_{\rm fd}n_fn_d,
\label{eq:Efn}
\end{equation}
\begin{equation}
E_{\mathrm{av}}(f^{n+1}\underline{L})
=
E_{\mathrm{av}}^{(0)}(f^{n+1}\underline{L})
+
U_{\rm fd}(n_f+1)n_d.
\label{eq:Efn1}
\end{equation}
The charge-transfer energy including the $f$--$d$ Coulomb contribution is then obtained as
\begin{equation}
\Delta'
\equiv
E_{\mathrm{av}}(f^{n+1}\underline{L})
-
E_{\mathrm{av}}(f^{n})
=
\Delta^{(0)}
+
U_{\rm fd}n_d.
\label{eq:Deltap}
\end{equation}

Experimentally, the charge-transfer energy determined from spectroscopy corresponds to $\Delta'$, which explicitly includes the $f$--$d$ Coulomb contribution within the present model. The present result identifies $\Delta'$ as the spectroscopic counterpart of the renormalized $f$-level energy appearing in the Falicov--Kimball and extended periodic Anderson models \cite{Falicov1969PRL, Watanabe2010PRL, Watanabe2009JPSJ, Watanabe2018IOP}, thereby establishing a direct connection between core-level spectroscopy and the low-energy parameter governing valence behavior.
We emphasize that $\Delta'$ does not introduce a new definition, but rather makes explicit the microscopic origin of the charge-transfer energy that has remained hidden in conventional phenomenological treatments.

\begin{figure*}[t]
\includegraphics[width=18.2cm]{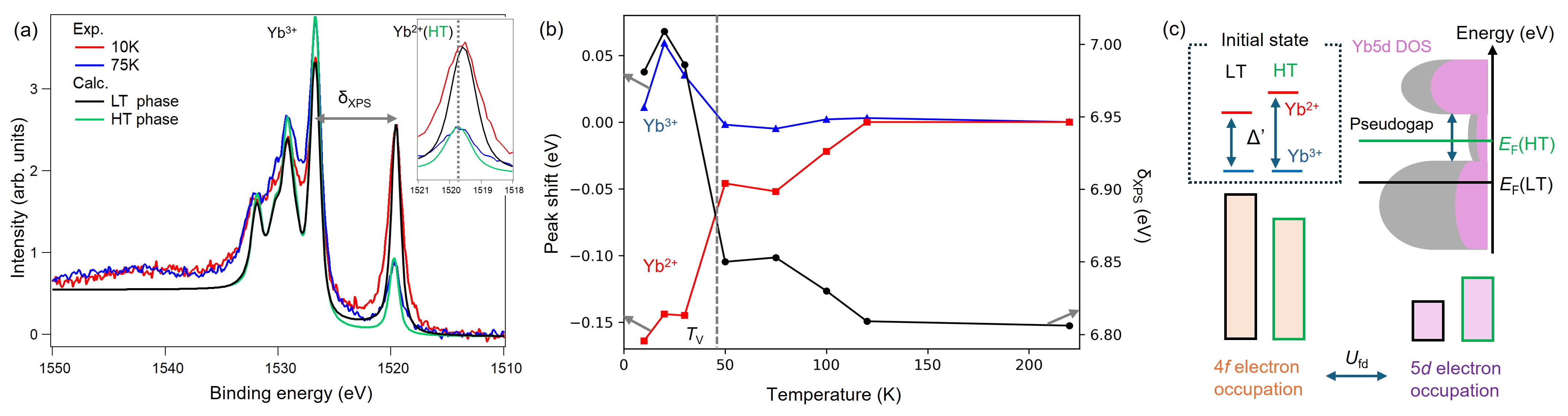}
\caption{
(Color online)
(a) Comparison between the calculated Yb $3d_{5/2}$ core-level XPS spectra for the low-temperature (LT, black) and high-temperature (HT, green) phases and the experimental spectra measured at 10 K (red) and 75 K (blue) \cite{Sato2004PRL}.
The calculated transition intensities were convoluted with a Gaussian broadening of 0.3~eV [half width at half maximum (HWHM)] to account for the experimental energy resolution and a Lorentzian broadening of 0.3~eV (HWHM) to account for the finite lifetime of the core-excited final states.
(b) Temperature dependence of the peak shifts of the Yb$^{2+}$ (red squares) and Yb$^{3+}$ (blue triangles) components (left axis). The peak positions are referenced to their values at 220 K, so that both shifts become zero at 220 K. Black circles show the experimental values of $\delta_{\mathrm{XPS}}$ defined in panel (a) (right axis), where the Yb$^{3+}$ peak lies at approximately 1527~eV.
(c) Schematic illustration of the initial state proposed in this work. The $4f$ and $5d$ electron occupations are coupled through the $f$--$d$ Coulomb interaction $U_{\rm fd}$. The gray region represents the occupied density of states below the Fermi level, while the purple region denotes the Yb $5d$ density of states. The occupied portion of the Yb $5d$ density of states corresponds to the $5d$ electron occupation $n_d$.
}
\end{figure*}

We here discuss the quantitative aspect of the $f$--$d$ Coulomb interaction term $U_{\rm fd} n_d$ appearing in Eq.~(\ref{eq:Deltap}). Importantly, the contribution of $U_{\rm fd}$ to the charge-transfer energy depends on the initial-state $5d$ occupation $n_d$, appearing through the term $U_{\rm fd}n_d$.
Therefore, in the limit of $n_d \sim 0$, one obtains $\Delta' \sim \Delta^{(0)}$, and the conventional description of the charge-transfer energy is naturally recovered.

While the number of $f$ electrons in rare-earth compounds has been extensively determined by core-level spectroscopies \cite{Sato2004PRL, YbAl3PFYXAS2008PRB, SatoPRB2017, Fujimori_2023}, an experimental estimate of the $d$-electron occupation remains challenging.
Although the local 5$d$ occupation can be estimated from first-principles band-structure calculations, its absolute value depends on the choice of orbital projection or spatial partitioning used to define the local 5$d$ states, making a unique quantitative assignment difficult \cite{Karp2021PRB}.
Nevertheless, the valence transition in YbInCu$_4$ can be discussed within the present framework based on the charge-transfer energy $\Delta'$. The key point is that our discussion focuses on the change in the $5d$ occupation across the first-order valence transition, rather than its absolute value. Accordingly, the relevant quantity in the present framework is $U_{\rm fd}\delta n_d$, where $\delta n_d$ denotes the difference in $d$-electron occupation between the competing phases.

Figure~2 summarizes the central role of the charge-transfer energy $\Delta'$ in describing the valence transition of YbInCu$_4$. Figure~2(a) compares the experimental and calculated Yb $3d_{5/2}$ core-level x-ray photoemission spectroscopy (XPS) spectra for the low-temperature (LT) and high-temperature (HT) phases.
The calculated spectra were obtained using the impurity Anderson model (IAM) \cite{Tonai2017JPSJ, Sasabe2017JPSJ, Fujimori_2023}.
We employed the configuration interaction between $\ket{f^{14}}$ and $\ket{f^{13}C}$ where $C$ denotes a conduction-band electron. The rectangular conduction-band ($W=2$ eV), the number of discretized conduction-band states ($N=50$), and the core-hole Coulomb interaction ($U_{\rm fc}=10.2$ eV) were fixed, whereas the charge-transfer energy $\Delta'$ and the $c$--$f$ hybridization strength $V_{\rm cf}$ were optimized to reproduce the experimental spectra.
The optimized parameter sets ($\Delta'$, $V_{\rm cf}$) for LT and HT phases were (0.6, 0.275) and (0.7, 0.24) eV, respectively.
Additionally, the hybridization strength was reduced to 80\% of its original value in the presence of a core hole \cite{RIXS2001PRB}.
The IAM calculations included full multiplet effects within the Hartree-Fock-Slater (HFS) method~\cite{cowan1981theory}. The Slater integrals were reduced to 80\% of their HFS values \cite{Ufc1992JPSJ, Ufc1993PRB, Ufc2002PRB}.
The calculation successfully reproduces the characteristic spectral evolution across the transition, including the energy shift of the Yb$^{2+}$ peak.
The optimized parameters also yield Yb valences of 2.803 and 2.924 for the LT and HT phases, respectively, in good agreement with previous experimental reports \cite{Cornelius1997PRB, Dallera2002PRL, Kotani2012EPJB}.
To examine this evolution in greater detail, Fig.~2(b) shows the temperature dependence of the peak shifts of the Yb$^{2+}$ and Yb$^{3+}$ components together with the energy difference, $\delta_{\rm XPS}$, defined in Fig.~2(a). Both the Yb$^{2+}$ peak shifts and $\delta_{\rm XPS}$ exhibit an abrupt change of approximately 0.15~eV around the valence-transition temperature, $T_{\mathrm{V}}=42$ K, demonstrating the first-order nature of the transition, $n_f({\rm LT})-n_f({\rm HT})\approx0.12$. The agreement between experiment and calculation is consistent with changes in the initial-state electronic structure across the transition.

The microscopic origin of this modification is illustrated schematically in Fig.~2(c). In the present framework, the $4f$ and $5d$ electron occupations are coupled through the interorbital Coulomb interaction $U_{\rm fd}$. Since the occupied portion of the Yb $5d$ density of states (DOS) determines the $5d$ electron occupation $n_d$, changes in the electronic structure near the Fermi level directly modify the charge-transfer energy according to Eq.~(\ref{eq:Deltap}). Previous studies have suggested that the Fermi level is located near the edge of the pseudogap in the LT phase, whereas it shifts into the pseudogap region in the HT phase \cite{UtsumiPRB2011, JarrigePRL2015}. This redistribution of the occupied Yb $5d$ states changes $n_d$, resulting in a corresponding variation of $\Delta'$. Therefore, the experimentally observed XPS peak shift supports the present picture that the charge-transfer energy is governed by the initial-state 5$d$ electron occupation through $U_{\rm fd}$.

\begin{figure*}[t]
\includegraphics[width=18.2cm]{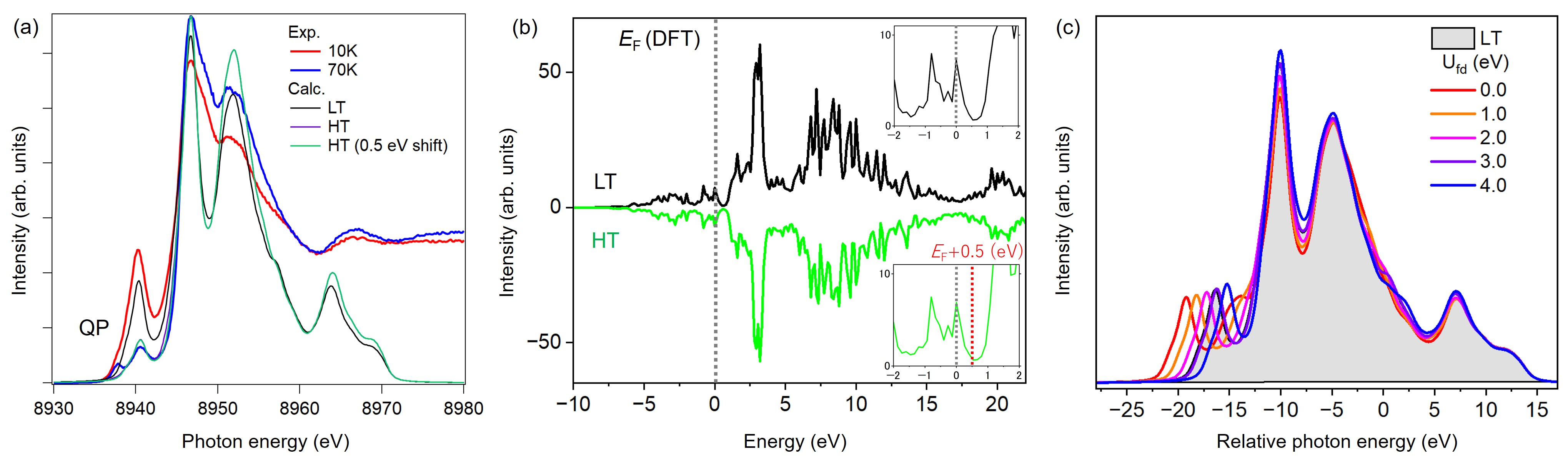}
\caption{
(Color online)
(a) Experimental and calculated HERFD-XAS spectra of YbInCu$_4$. The red and blue curves show the experimental spectra measured at 10~K and 70~K, respectively.
The overall experimental energy resolution was estimated to be approximately 0.68 eV (full width at half maximum).
The black and purple curves correspond to the calculated low-temperature (LT) and high-temperature (HT) phases, respectively. The green curve represents the HT spectrum calculated with an illustrative 0.5~eV shift of the Fermi level. QP denotes the quadrupole transition. The calculated transition intensities were convoluted with a Gaussian broadening of 0.3~eV [half width at half maximum (HWHM)] to account for the experimental energy resolution and a Lorentzian broadening of 0.5~eV (HWHM) for the finite lifetime of the core-excited final states.
(b) Calculated Yb $5d$ partial density of states (DOS) for the LT and HT phases. The gray dotted line indicates the Fermi level obtained from the DFT calculations. The insets enlarge the DOS near the Fermi level, highlighting the pseudogap structure.
The red dashed line in the HT inset indicates the illustrative Fermi-level position shifted by 0.5~eV relative to the DFT value, as used for the green spectrum in Fig.~3(a).
(c) Calculated HERFD-XAS spectra for different values of $U_{\rm fd}$,
with all other parameters fixed to those used for the LT phase in panel (a). The gray shaded area represents the calculated LT spectrum obtained with $U_{\rm fd}$=2.8~eV.
}
\end{figure*}

We next investigate the microscopic origin of the first-order valence transition by analyzing the $L_{\rm III}$ ($2p\rightarrow5d$) high energy resolution fluorescence detected x-ray absorption spectroscopy (HERFD-XAS) spectra \cite{Kawamura2017JPSJ}.
In the XPS analysis, $U_{\rm fd}$ is incorporated into the charge-transfer energy $\Delta'$ and cannot be determined independently. In contrast, as illustrated in the right panel of Fig.~1, HERFD-XAS is directly sensitive to $U_{\rm fd}$ through the $2p\rightarrow5d$ excitation, allowing its magnitude to be estimated from the experimental spectra.

Figure~3(a) compares the experimental and calculated HERFD-XAS spectra for the low-temperature (LT) and high-temperature (HT) phases. Using the corresponding charge-transfer energy $\Delta'$ and $c$--$f$ hybridization strength $V_{\rm cf}$ determined from the XPS analysis for each phase, the calculations reproduce the spectral evolution across the transition, including the relative energy positions of the Yb$^{2+}$ (8941 eV) and Yb$^{3+}$ (8947 eV) features and the small energy shift of the Yb$^{2+}$ peak. As illustrated in Fig.~1, the additional final-state contribution of $U_{\rm fd}$ is required to reproduce the relative spectral energies. By optimizing $U_{\rm fd}$ to reproduce the experimental HERFD-XAS spectra, we obtain $U_{\rm fd}\approx2.8$ eV.

The calculations are based on the Yb $5d$ DOS obtained from first-principles calculations using WIEN2k, where the small structural change across the valence transition, $I\bar{4}m2$ (\#119, LT phase) $\rightarrow$ $F\bar{4}3m$ (\#216, HT phase), is taken into account \cite{Tsutsui2016JPSJ}, as shown in Fig.~3(b).
The calculated Yb $5d$ DOS is nearly identical in the LT and HT phases, with a pseudogap structure present just above the Fermi level in both phases. Previous experimental studies suggest that the Fermi level moves into the pseudogap region in the HT phase \cite{JarrigePRL2015, UtsumiPRB2011}, as schematically illustrated in Fig.~2(c). However, such a change is not clearly reproduced by the present band-structure calculations, possibly reflecting many-body effects associated with valence fluctuations.
To examine whether such a Fermi-level shift can be detected in HERFD-XAS, we calculated the spectra using both the unshifted DOS (purple) and an illustrative 0.5 eV shift that places the Fermi level near the center of the pseudogap (green). As shown in Fig.~3(a), the resulting spectral differences are minor. In addition, the near-Fermi-level Yb $5d$ states overlap with the quadrupole ($2p\rightarrow4f$) transition, further obscuring the corresponding spectral changes. Thus, although the Fermi-level shift is difficult to resolve directly in HERFD-XAS, the pseudogap-related electronic reconstruction suggested by previous experiments is consistent with the physical picture proposed in this study.

Figure~3(c) shows the influence of $U_{\rm fd}$ on the HERFD-XAS spectra. While the overall spectral shape remains nearly unchanged, the Yb$^{2+}$ peak systematically shifts toward higher energy with increasing $U_{\rm fd}$. The best agreement with the experimental spectrum is obtained for $U_{\rm fd}\approx2.8$~eV.
Based on the above spectroscopic analyses, we summarize the electronic behavior associated with the first-order valence transition in YbInCu$_4$. From the Yb $3d$ XPS results, the charge-transfer energy in the HT phase is larger than that in the LT phase by approximately 0.1 eV, i.e.,
$\delta\Delta=\Delta'({\rm HT})-\Delta'({\rm LT})\approx0.1$ eV, consistent with the HERFD-XAS analysis. The HERFD-XAS spectra further yield
$U_{\rm fd}\approx2.8$ eV. Assuming that $\Delta^{(0)}$ and $U_{\rm fd}$ remain constant across the transition, we estimate the change in the $5d$-electron occupation to be
$\delta n_d=n_d({\rm HT})-n_d({\rm LT})\approx0.04$.
This value is consistent with the change in the $4f$ occupation,
$\delta n_f=n_f({\rm LT})-n_f({\rm HT})\approx0.12$,
indicating a reasonable redistribution of charge between the $4f$ and $5d$ states across the transition.

These results establish a microscopic connection between the charge-transfer energy determined from core-level spectroscopy and the $f$--$d$ Coulomb interaction central to theoretical models describing low-energy valence instability in rare-earth compounds.
On the other hand, previous studies have emphasized the roles of $c$--$f$ hybridization and conduction-band reconstruction in the valence transition of YbInCu$_4$ \cite{Kotani2012EPJB, JarrigePRL2015}.
The reconstruction of the electronic states near the pseudogap can affect both the effective hybridization and the Yb $5d$ electron occupation across the transition.
In principle, similar XPS spectral changes can also be reproduced by varying $V_{\rm cf}$ alone, depending on the choice of $\Delta^{(0)}$ and $U_{\rm fc}$.
However, our HERFD-XAS analysis reveals a substantial $f$--$d$ Coulomb interaction of $U_{\rm fd}\approx 2.8$ eV, which cannot be neglected in describing the initial-state electronic structure.
Even without a drastic change in $c$--$f$ hybridization \cite{DMFT2023NCR}, changes in the $5d$ occupation associated with conduction-band reconstruction may affect the charge-transfer energy through $U_{\rm fd}n_d$.
Therefore, both the charge-transfer energy, including the $f$--$d$ Coulomb contribution, and $c$--$f$ hybridization should be considered on an equal footing.

In this work, we established a spectroscopic framework demonstrating that, in rare-earth compounds, the conventionally defined charge-transfer energy, denoted by $\Delta^{(0)}$, acquires an additional $f$--$d$ Coulomb contribution, $U_{\rm fd}n_d$, arising from the initial-state electronic configuration. The resulting charge-transfer energy is expressed as $\Delta'=\Delta^{(0)}+U_{\rm fd}n_d$.
This formulation provides a microscopic connection between high-energy spectroscopic observables and low-energy electronic properties, extending the conventional phenomenological interpretation of the charge-transfer energy in correlated $f$-electron systems.
Importantly, the $U_{\rm fd}n_d$ contribution is not restricted to changes across first-order valence transitions. The initial-state $5d$ occupation $n_d$ contributes to the energy separation between different $f$-electron configurations through the $f$--$d$ Coulomb interaction, even in systems without pronounced valence changes. Thus, the present formulation provides a microscopic interpretation of the phenomenological $f$-level energies commonly employed in photoemission models.

Applying this framework to the mixed-valence compound YbInCu$_4$, we demonstrated that the spectroscopic evolution across the first-order valence transition can be consistently described in terms of changes in the charge-transfer energy. Combined analyses of the 3$d$ XPS and $L_{\rm III}$ HERFD-XAS reveal that the evolution of $\Delta'$ is governed by changes in the initial-state $5d$-electron occupation associated with the pseudogap near the Fermi level. Furthermore, the HERFD-XAS analysis yields $U_{\rm fd}\approx2.8$ eV, enabling a quantitative estimate of the corresponding change in the $5d$-electron occupation across the transition.

The present formulation is not merely a reparameterization of the conventional charge-transfer energy, but reveals that the spectroscopically determined charge-transfer energy inherently contains a many-body contribution originating from the $f$--$d$ Coulomb interaction. This framework provides a unified microscopic perspective for correlated rare-earth compounds exhibiting valence instability \cite{HAXPES2007PRB, Watanabe2013JPSJ, PrSciAd2024}, ranging from first-order valence-transition systems, such as metallic Ce undergoing the $\gamma$--$\alpha$ transition \cite{Koskenmaki1978, WuilloudPRB1983, TegomePRB2023}, SmS \cite{Coey1976PRB, Deen2005PRB, Watanabe2021JPSJ}, and EuPd$_2$Si$_2$ \cite{Bauminger1973PRL, Wada1996JPSJ, Mimura_2011}, to valence-fluctuation quantum-critical materials including $\beta$-YbAlB$_4$ \cite{bYbAlB4_2008NatPhy, Kuga2008PRL, YbAlB4_ESR_2011_PRL} and CeCu$_2$Si$_2$ \cite{Steglich1979PRL, Holmes2004PRB, Seyfarth2012PRB}.
The present framework provides a basis for understanding quantum valence criticality and related emergent phenomena in correlated $f$-electron materials \cite{Miyake2014JPSJ}.

\begin{acknowledgments}
We are grateful to H. Fujiwara and K. Kuga for useful discussions. The experiments were performed at SPring-8 with the approval of JASRI (Proposal Nos. 2017B1223, 2019B1599, 2020A2142, 2021B1534, 2022A1622, and 2023A1242). This work was partly supported by JSPS KAKENHI (Grant Nos. 21H03739, 21K03467, and 23H04870) and JST CREST (Grant No. JPMJCR2435).
\end{acknowledgments}

\bibliographystyle{apsrev4-2}
\bibliography{bib}

\end{document}